# Enhanced Breakdown and RF Performance in Field-Plated AlGaN/GaN HEMT for High-Power Applications

Tanjim Rahman, Trupti Ranjan Lenka

Department of Electrical and Computer Engineering, Texas Tech University, Lubbock, TX, 79409, USA

Emails: tanjrahm@ttu.edu, tlenka@ttu.edu

**Abstract**

High Electron Mobility Transistors (HEMTs) are most suitable for harsh environments as they operate reliably under extreme conditions such as high voltages, high temperatures, radiation exposure and corrosive atmospheres. In this article gate field-plated engineering $Al_{0.295}GaN/GaN$ HEMT is proposed for achieving high breakdown voltage to reliably operate in harsh environments. The $Al_{0.295}GaN/GaN$ heterointerface results in a 2DEG (two-dimensional electron gas) density of the order of $10^{13}$ cm$^{-2}$ obtained from the self-consistent solution of Schrödinger and Poisson equations. The device has undergone DC and breakdown simulations which result in threshold voltage of -5.5 V, drain saturation current of 3000 mA and breakdown voltage of 1 kV. The HEMT also shows excellent RF characteristics which include cut-off frequency ($f_t$) of 28 GHz and maximum frequency of oscillation ($f_{max}$) of 38 GHz. The proposed gate field-plated HEMT is stable up to 40 GHz and suitable for high-voltage, and high-power RF operation during harsh environment applications.

*Key Words: AlGaN, Breakdown, Field-Plate, GaN, HEMT*

## I. Introduction

GaN is the preferred material of worldwide researchers due to its unique material properties for electronic device applications. The material properties include wide bandgap (~3.4 eV), direct bandgap semiconductors, high polarization, high electron mobility, high saturation velocity (~$10^7$ cm/s), high electric field (~6 MV/cm) and high thermal conductivity[1]. Two types of polarizations exist, *i.e.* spontaneous and piezoelectric polarization in III-nitrides[2]. The polarization property of $Al_xGa_{1-x}N$, when epitaxially grown over GaN induces two-dimensional electron gas (2DEG) at the heterointerface which is referred to as channel[3]. The mobility of electrons becomes very high due to lack of scattering in this channel hence high electron mobility transistor action is achieved. The high electron mobility results in faster switching and lower resistance[4]. High 2DEG density leads to high current and high frequency of oscillations. AlGaN/GaN HEMT is a type of field-effect transistor (FET) that's best suitable for high-power and high-frequency applications[5]. Unlike Silicon based FETs, it can handle higher voltages and temperatures and is most suitable for harsh environments.

The key advantages of AlGaN/GaN HEMT include high breakdown voltage which can handle large voltages, high efficiency which enables lower losses in power conversion, fast switching speeds which are ideal for RF/microwave applications and high temperature tolerance which enables more robust operation in harsh environments[6], [7]. However, in high-voltage operation, AlGaN/GaN HEMTs suffer from very high electric field concentration near the drain-edge of the gate which limits breakdown voltage, cause reliability issues like trapping effects and reduce lifetime [8]. Whereas field-plate technology helps to spread the electric field more evenly across the device which helps increase the breakdown voltage, improve reliability by suppressing current collapse which is a trapping-related performance degradation effect and better RF power handling capability [9], [10]. Field plate length and position impact electric field control and capacitance, where there is a trade-off between increased capacitance ($C_{gd}$) and switching speed[8]. In field-plate technology, the role of passivation layer and its quality is crucial to prevent leakage or trapping effects [9], [11].

The key advantages of field-plated AlGaN/GaN HEMT include high breakdown voltage, reduced electric field crowding, lower current collapse, enhanced RF performance, better power handling capability, and improved device reliability[9], [12]. Hence in this research, to increase the breakdown voltage up to 1 kV and to reliably operate in harsh environment conditions, gate field-plate engineered AlGaN/GaN HEMT is proposed.

The state-of-the-art research on field-plate technology such as a Schottky gate on p-GaN and a source field-plate are used to achieve a positive threshold voltage and optimized breakdown performance, reaching 725 V and 239 V at a field-plate length ($L_{GD}$) of 7 µm, respectively [13]. Various field-plate designs in AlGaN/GaN HEMTs were investigated using TCAD and fabrication, confirming that the gate-source composite field plate achieved the highest breakdown voltage and a figure of merit (FOM) of 504 MW/cm² by enhancing breakdown and reducing on-resistance [14]. Breakdown voltage of HEMTs was analyzed for different field-plate lengths ($L_{FP}$ = 1 to 2 µm), showing a peak $V_{BR}$ of 871V and JFoM of 34.88 THz·V at $L_{FP}$ = 1.75 µm, which was further increased to 912 V with a high-k $HfO_2$ passivation layer [15], [16]. The impact of drain field-plate length on semi-ON degradation in AlGaN/GaN HEMTs was analyzed, showing that longer drain FPs lead to faster drain current degradation due to enhanced hot electron trapping at the passivation/barrier interface from altered electric field distribution[17]. To suppress high electric field in the SEMI-ON state, a 30 nm $SiO_2$ pocket was incorporated at the field plate edge, resulting in a 43% reduction in electric field, 20% reduction in electron temperature, 13% decrease in self-heating, and 47% drop in thermal resistance[16]. RF/DC performance of 250 nm gate AlGaN/GaN HEMTs on silicon showed that with field-plate lengths of 0.5 µm, 1 µm, and 2 µm, the devices achieved breakdown voltage/cut-off frequency ($f_T$) of 200 V/145 GHz, 370 V/125 GHz, and 440 V/100 GHz, respectively, while maintaining high power and RF performance for future applications [18]. An embedded metal layer was proposed to suppress strong electric fields at the field-plate edge in AlGaN/GaN HEMTs, leading to ~3% reduction in electric field, 20% decrease in converse piezoelectric effect, and 14% drop in electron temperature, improving reliability[19]. A high-performance E-mode GaN HEMT with a source-connected field-plate achieved a threshold voltage of 2.81 V, maximum current of 757 mA/mm, and breakdown voltage of 866 V[20]. The RF performance of AlGaN/GaN HEMTs was further enhanced by introducing an air-bridged source-connected field-plate with asymmetric passivation, resulting in a 31.6% increase in cut-off frequency and 7.8% rise in $f_{max}$, with only 1.4% error compared to measured data and no degradation in DC performance [21]. TCAD evaluation of a source-connected field-plate GaN HEMT showed improved OFF-state breakdown characteristics for mm-wave and space applications, with reliability analysis under heavy ion strike[22]. A fast optimization method for gate-source dual field plates in AlGaN/GaN HEMTs using an artificial neural network and particle swarm optimization achieved a breakdown voltage of 1228 V with only 3.06% prediction error, enabling automated and efficient field plate design[23]. A 60-nm gate-length graded-channel AlGaN/GaN HEMT with a mini-field-plate T-gate achieved $f_T$ of 156 GHz, $f_{MAX}$ of 308 GHz, 5.5 W/mm output at 30 GHz, and a record $f_T$ and power density product of 858 GHz·W/mm with minimal current collapse[24]. A grated gate field-plate HEMT was proposed to improve conventional gate field-plate designs, achieving 1A/mm output current, 350mS/mm transconductance, and a 60% increase in cut-off frequency (28.3 GHz vs. 17.6 GHz), by effectively reducing parasitic capacitances [25].

The design of proposed gate field-plated $Al_{0.295}GaN/GaN$ HEMT and its model development is thoroughly presented in Section II. The DC characteristics and Breakdown voltage are discussed in Section III and IV respectively. The RF characteristics of the proposed structure are presented in Section V. Finally, the research is concluded in Section VI.

## II. Device Structure and Model Development

The proposed gate field-plated $Al_{0.295}GaN/GaN$ HEMT is shown in Figure 1. The device structural parameters are listed in Table 1. The device has been simulated using Silvaco TCAD to understand its electrical characteristics for potential applications in high-voltage, high-power and high-frequency operations.

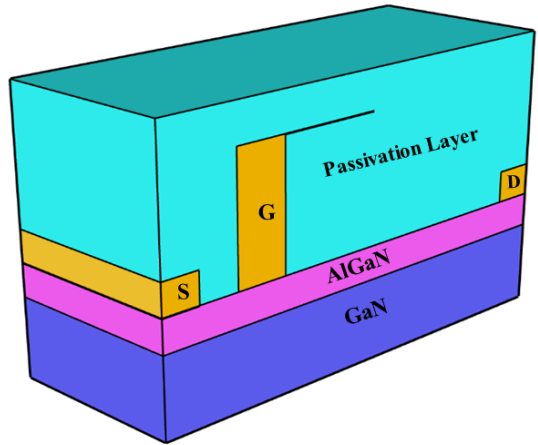

Fig 1: Cross-sectional View of the Proposed Gate Field-Plated $Al_{0.295}GaN/GaN$ HEMT Structure

The structural design of the proposed gate field-plated $Al_{0.295}GaN/GaN$ HEMT consists of a 0.7 µm gate length ($L_g$) with a vertical gate height of 1.5 µm and a gate metal work function of 5.23 eV. The device incorporates a gate field plate extending 1 to 2 µm horizontally toward the drain, with a gate-to-drain spacing ($L_{gd}$) of 5.0 µm to support high breakdown performance. A passivation layer with a thickness of 0.4 µm covers the device surface and field plate to suppress leakage and trapping effects. The epitaxial stack includes a 30 nm AlGaN barrier layer for polarization-induced 2DEG formation and a 180 nm undoped GaN channel layer beneath it. Doping profiles include a Gaussian n-type implant of $1\times10^{18}$ cm$^{-3}$ for the source/drain, uniform n-type doping of $1\times10^{16}$ cm$^{-3}$ in the AlGaN barrier, and $1\times10^{15}$ cm$^{-3}$ in the GaN channel. Carrier lifetime values for both electrons and holes ($\tau_n$ and $\tau_p$) are set to $1\times10^{-9}$ s, modeled using the Shockley-Read-Hall (SRH) recombination approach.

In simulating gate field-plated $Al_{0.295}GaN/GaN$ HEMTs, several key physical models have been used to capture the complex behaviors of the proposed device, especially the effects of field-plates. The models used are carrier transport models which govern how electrons move through the device such as drift-diffusion (DD) model for DC and low frequency operation, and hydrodynamic (HD) model which captures hot-carrier effects with carrier energy balance and velocity overshoot, which is useful for high-power and high-frequency behaviors of AlGaN/GaN HEMTs. In addition to Poisson and continuity equations, HD solves the carrier energy balance as per Eq (1) as follows.

$$\frac{3}{2}k\frac{\partial(nT_n)}{\partial t} + \nabla \cdot S_n = J_n \cdot E - R_n E_n \qquad (1)$$

where $T_n$ is the electron temperature, $S_n$ is energy flux vector, $J_n$ is electron current density, and $E_n$ electron energy relaxation time. The current density equations are modified to include temperature effects as per Eq (2) as follows

$$J_n = qn\mu_n E + qD_n \nabla n + qn\mu_n \nabla T_n \qquad (2)$$

To handle self-heating effects in AlGaN/GaN heterostructure, self-heating models have been invoked in the simulation process. The polarization models include spontaneous and piezoelectric polarization which are critical for AlGaN/GaN heterostructures [2], [26]. These models generate 2DEG at the AlGaN/GaN interface as shown in Figure 2. To accurately model 2DEG confinement under the AlGaN/GaN heterointerface density gradient model has been enabled. The field-plate related effects have been considered imposing interface trap models and dielectric models in the simulation. The electron mobility is strongly field-dependent and temperature-dependent. Hence, a high-field mobility saturation model is used. SRH (Shockley-Read-Hall) and Auger recombination models have been used for carrier recombination. The SRH recombination rate ($R_{SRH}$) explains carrier recombination in AlGaN/GaN via trap states (defects or impurities) within the bandgap which is expressed in Eq (3) as follows.

$$R_{SRH} = \frac{np - n_i^2}{\tau_p(n+n_1) + \tau_n(p+p_1)} \qquad (3)$$

where, n and p are electron and hole concentrations, $n_i$ is the intrinsic carrier concentration, $\tau_n$, and $\tau_p$ are electron and hole lifetimes.

The high-field saturation mobility model explains how mobility of carriers decreases at increased electric field, due to velocity saturation effects in AlGaN/GaN interface which is expressed in Eq (4) as follows.

$$\mu(E) = \frac{\mu_0}{\left[1+\left(\frac{\mu_0 E}{v_{sat}}\right)^{\frac{1}{\beta}}\right]^{\beta}} \qquad (4)$$

where $\mu_0$ is the low-field mobility, $v_{sat}$ is the saturation velocity (~ $2 \times 10^7$ cm/s for GaN), E is the lateral electric field and $\beta$ is empirical fitting parameter (1 to 2).

The Schrödinger and Poisson equations were solved self-consistently at the AlGaN/GaN heterointerface to model 2DEG formation and subband structure [27]–[29]. As shown in Figures 2–4, the simulation reveals a 2DEG forming ~30 nm below the surface, with corresponding electric field distribution and polarization-induced electron concentration.

Peak electric fields of 1.5 MV/cm and a polarization-induced electron concentration of $8 \times 10^{19}$ cm$^{-3}$ are observed at the AlGaN/GaN heterointerface in Figures 3 and Figures 4.

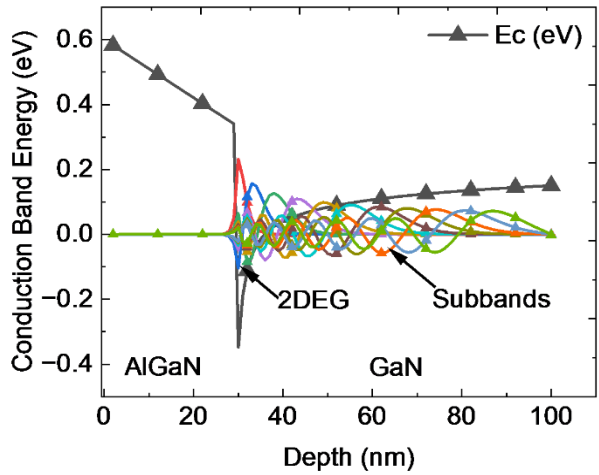

Fig 2: Self-Consistent 2DEG Formation at the $Al_{0.295}GaN/GaN$ heterointerface

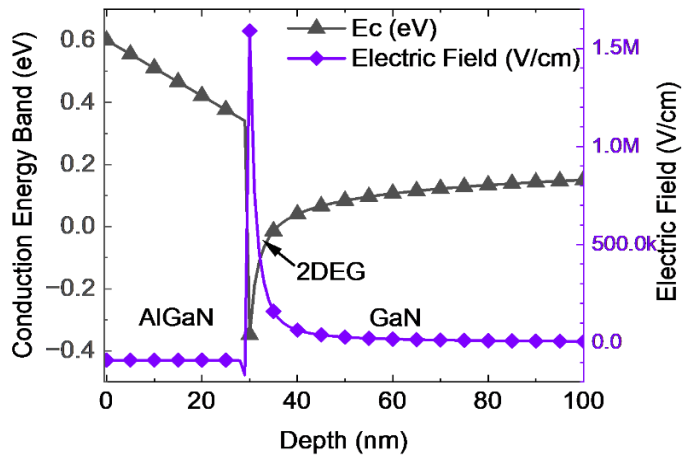

Fig 3: Realization of Electric Field in the 2DEG of Gate Field-Plated $Al_{0.295}GaN/GaN$ HEMT

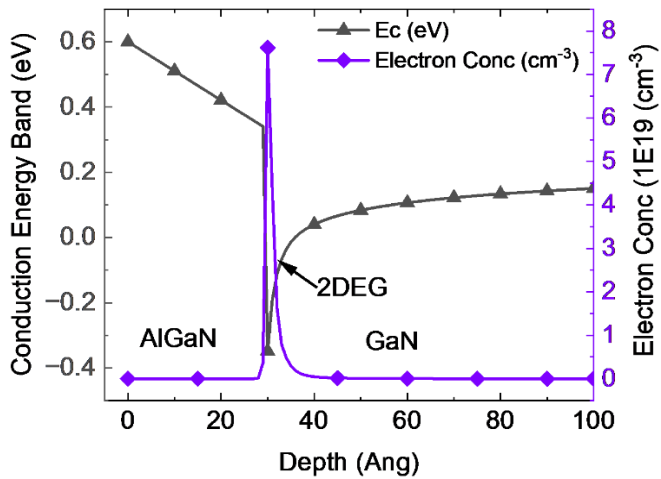

Fig 4: Polarization induced Electrons in the 2DEG of Gate Field-Plated $Al_{0.295}GaN/GaN$ HEMT

## III. DC Characteristics

The DC characteristics of the proposed gate field-plated AlGaN/GaN HEMT have been analyzed through input transfer characteristics and output drain characteristics. The HEMT is simulated with a gate bias sweeping from -8V to 0V with a constant drain bias of 1V as shown in Figure 5. The drain current increases beyond −5.5 V and peaks at 12 mA at VGS = 0 V, confirming normally-on (depletion-mode) operation with a threshold voltage of −5.5 V. The peak transconductance is 12 mS, as shown in Figure 5.

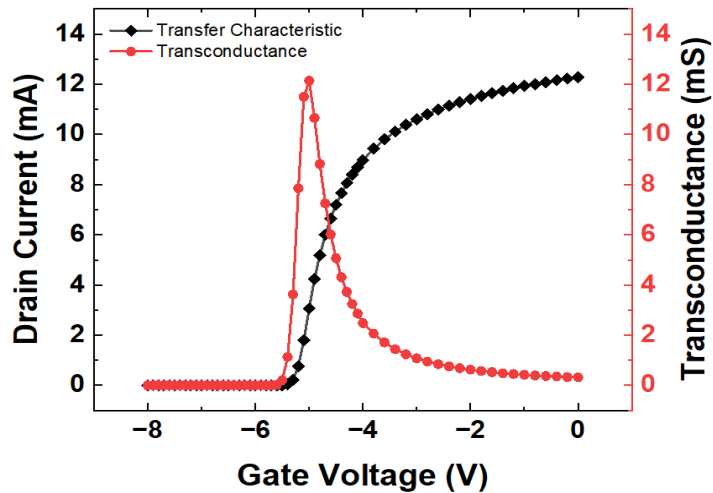

Fig 5: Transfer Characteristics ($I_{ds}$-$V_{gs}$) of Gate Field-Plated $Al_{0.295}$GaN/GaN HEMT

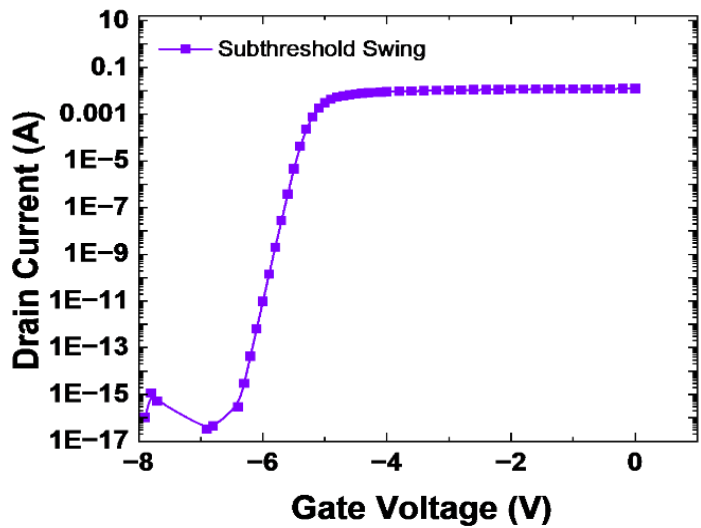

Fig 6: Subthreshold Slope of Gate Field-Plated $Al_{0.295}$GaN/GaN HEMT

The subthreshold swing (SS) of the proposed HEMT is 125 mV/dec in Figure 6, indicating good switching performance and low leakage. This value, typical for GaN HEMTs, exceeds the ideal MOSFET limit (~60 mV/dec) due to differing transport mechanisms and device structure.

$$SS = \left(\frac{dV_{gs}}{d\log I_d}\right) \tag{5}$$

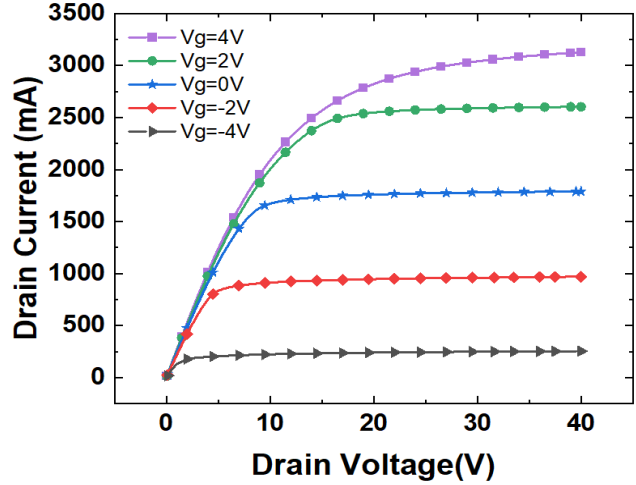

Fig 7: Drain Characteristics (Ids-Vds) of Gate Field-Plated Al$_{0.295}$GaN/GaN HEMT

The drain characteristics of the AlGaN/GaN HEMT, shown in Figure 7, were analyzed for gate voltages ranging from –4 V to 4 V. The device exhibits typical linear and saturation regions, with a peak drain current of 3000 mA at 4 V gate bias and 40 V drain bias, confirming its suitability for high-power electronic applications.

## IV. Breakdown Voltage Characteristics

The gate field-plated AlGaN/GaN HEMT was simulated in TCAD to analyze breakdown behavior under various model parameters. The electric field distribution is shown in Figure 8, and breakdown characteristics were evaluated for different passivation layers HfO$_2$, Al$_2$O$_3$ and Si$_3$N$_4$ as illustrated in Figure 9, Figure 10 and Figure 11.

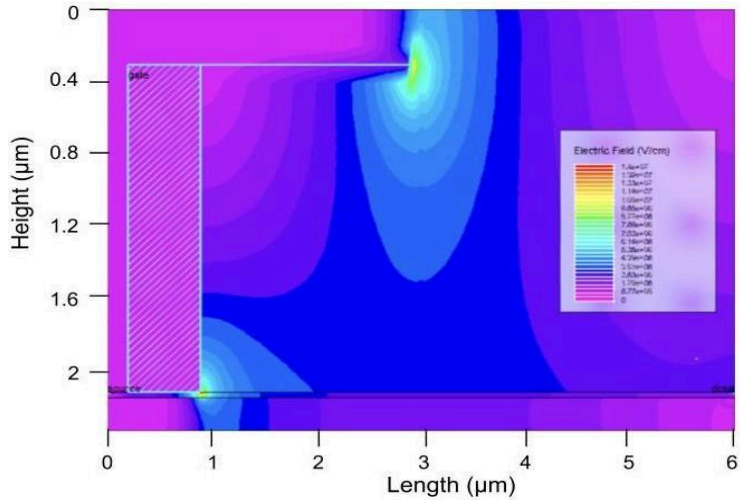

Fig 8: Simulated electric field distribution in the Gate Field-Plated Al$_{0.295}$GaN/GaN HEMT

The breakdown voltage is analyzed at different field-plate lengths (L) from 1 μm to 2 μm with a step of 0.2 μm. It is observed that the device with gate field-plate length of 2 μm breaks down at 1 kV when $HfO_2$ dielectric is used and at different field plate lengths all curves follow the same trend. The breakdown voltage characteristics of gate field-plated AlGaN/GaN HEMT with $Al_2O_3$ passivation layer is less than 1 kV due to less dielectric constants which are demonstrated in Figure 10.

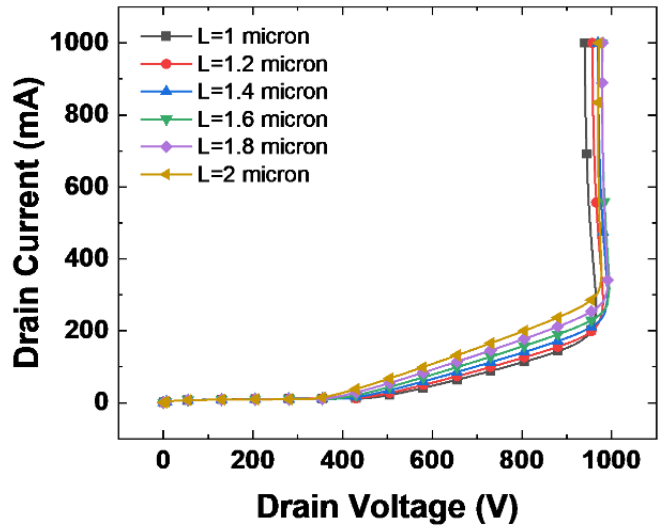

Fig 9: Breakdown Voltage of Gate Field-Plated $Al_{0.295}GaN/GaN$ HEMT with $HfO_2$ Passivation Layer

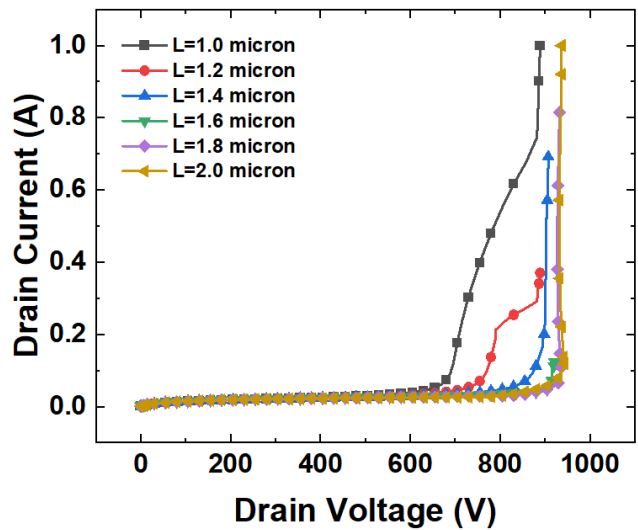

Fig 10: Breakdown Voltage of Gate Field-Plated $Al_{0.295}GaN/GaN$ HEMT with $Al_2O_3$ Passivation Layer

The breakdown voltage is also analyzed at different field-plate lengths (L) from 1 μm to 2 μm with a step of 0.2 μm for $Si_3N_4$ passivation layer which is shown in Figure 11. It is observed that the HEMT breaks down early prior to 1 kV. However, all the curves follow the same trends. In all the three cases, it is observed that the HEMT with gate field-plate length of 2 μm, the device breaks down late.

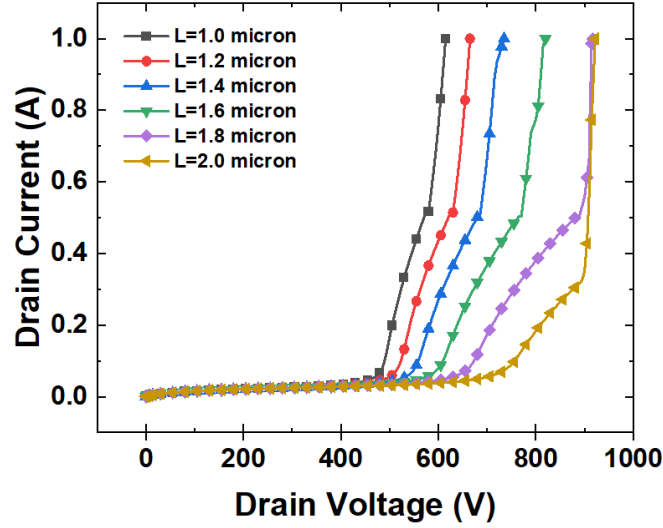

Fig 11: Breakdown Voltage of Gate Field-Plated Al$_{0.295}$GaN/GaN HEMT with Si$_3$N$_4$ Passivation Layer

## V. RF Characteristics

AC simulations of the gate field-plated AlGaN/GaN HEMT yield RF parameters including ft, fmax, Gma, and Gms [30]–[32]. Cut-off frequency is the transition frequency where the current gain becomes unity, and it gives a measure of how much faster a transistor can operate. High electron mobility of HEMT, short gate lengths and low parasitic capacitance are the influencing factors of cut-off frequency. It is proportional to transconductance ($g_m$) and inversely proportional to gate to source ($C_{gs}$) and gate to drain capacitance ($C_{ds}$) as mentioned in Eq (6) as follows.

$$f_t = \frac{g_m}{2\pi(C_{gs} + C_{gd})} \quad (6)$$

The maximum frequency of oscillation (f$_{max}$) is the highest frequency at which a HEMT can sustain power gain of unity. It's one of the most critical high-frequency figures of merit (FOM), important especially in RF/microwave monolithic integrated circuit (MMIC) design applications. The maximum frequency of oscillation (f$_{max}$) and cut-off frequency (f$_t$) are related along with feedback miller capacitances ($C_{gd}$), gate ($R_g$) and output ($R_{out}$) resistances, and other parasitic effects as mentioned in Eq (7) as follows.

$$f_{max} \approx \frac{f_t}{2}\sqrt{\frac{g_m R_g}{C_{gd} R_{out}}} \quad (7)$$

The cut-off frequency (f$_t$) of 28 GHz is obtained from the current gain at 0 dB (unity) whereas maximum frequency of oscillation (f$_{max}$) of 38 GHz is obtained from unilateral power gain at 0 dB (unity) which is demonstrated in Figure 12.

The maximum available gain (Gma) and maximum stable gain (Gms) are related to the stability factor as mentioned in Eq (8) as follows.

$$Gma = \left|\frac{S_{21}}{S_{12}}\right|\left(K - \sqrt{K^2 - 1}\right) = Gms\left(K - \sqrt{K^2 - 1}\right) \quad (8)$$

Where $S_{21}$ and $S_{12}$ are scattering matrix parameters, K is the stability factor (K>1 for unconditionally stable and K<1 potentially unstable).

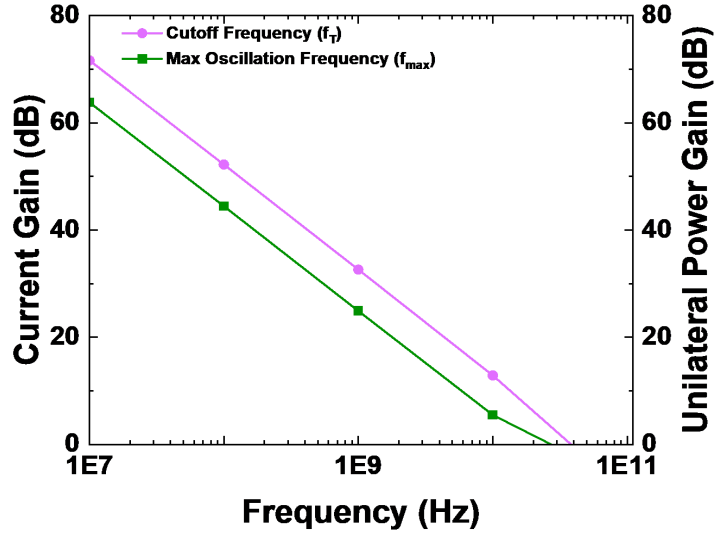

Fig 12: Cut-off Frequency and Max Frequency of Oscillation of Gate Field-Plated $Al_{0.295}GaN/GaN$ HEMT

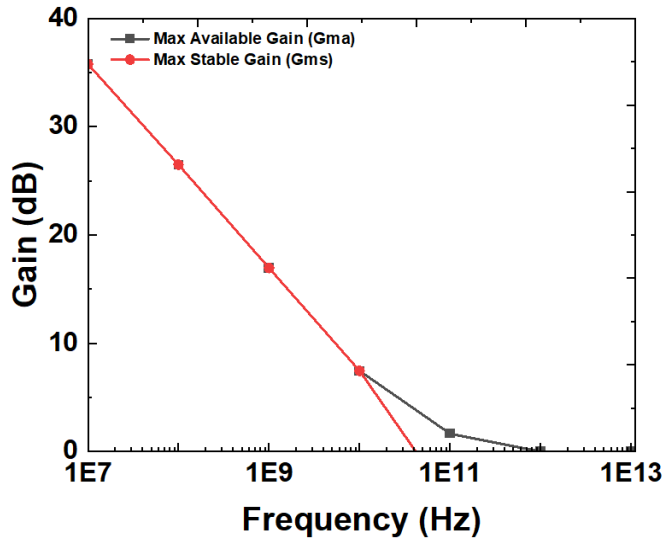

Fig 13: Maximum Available Gain, and Maximum Stable Gain of Gate Field-Plated $Al_{0.295}GaN/GaN$ HEMT

The maximum available gain (Gma) and maximum stable gain (Gms) of the gate field-plated $Al_{0.295}GaN/GaN$ HEMT, shown in Figure 13, exhibit a strong overlap from 10 MHz to 10 GHz, decreasing from 35 dB to 8 dB. This overlap indicates excellent stability, with the device remaining stable up to 40 GHz.

## VI. Conclusion

In this article gate field-plated $Al_{0.295}GaN/GaN$ HEMT is proposed, and field-plate engineering has been applied from L=1 μm to 2 μm to investigate the device performance and find the device breakdown voltage at different dielectric passivation layers. The device has undergone DC simulations which result in threshold voltage of -5.5 V and drain saturation current of 3000 mA. The subthreshold slope is obtained as 125 mV/decade. The breakdown voltage of 1 kV is obtained at gate field-plate length of 2 μm when $HfO_2$ passivation was used. The HEMT also shows excellent RF characteristics which comprise of cut-off frequency ($f_t$) of 28 GHz and maximum frequency of oscillation ($f_{max}$) of 38 GHz. The proposed HEMT is stable up to 40 GHz and very suitable for high-voltage and high-power RF operation in harsh environments.

## Acknowledgements

The authors acknowledge the Department of Electrical and Computer Engineering, Texas Tech University, Lubbock, TX, USA for providing necessary facilities to carry out the research. The author would also like to extend their sincere thanks to the Ultra Wide Bandgap Laboratory for providing advanced experimental and simulation tools. Special appreciation is due to Dr. Hieu P. Nguyen for his valuable guidance, mentorship and technical support throughout this course of this work.

## Data Availability Statement

All data that supports the findings of this study are included within the article.